\documentclass
[floatfix,superscriptaddress,secnumarabic,amssymb,amsmath,nobibnotes,aps,prd,showkeys,showpacs,nofootinbib,twocolumn,notitlepage,10pt]{revtex4}%
\usepackage{setspace}
\usepackage{color}
\usepackage{amsmath}
\usepackage{amsfonts}
\usepackage{verbatim}
\usepackage{amssymb}
\usepackage{graphicx,bm}
\usepackage{graphicx}
\usepackage{amsmath}
\usepackage{amssymb}
\usepackage{amssymb}
\usepackage{graphicx,bm}
\usepackage{graphicx}
\usepackage[caption=false]{subfig}%
\providecommand{\U}[1]{\protect\rule{.1in}{.1in}}

\newcommand{\be}{\begin{equation}}
\newcommand{\ee}{\end{equation}}

\newcommand{\mincir}{\raise
-3.truept\hbox{\rlap{\hbox{$\sim$}}\raise4.truept\hbox{$<$}\ }}
\newcommand{\magcir}{\raise
-3.truept\hbox{\rlap{\hbox{$\sim$}}\raise4.truept\hbox{$>$}\ }}

\ifx\pdfoutput\relax\let\pdfoutput=\undefined\fi
\newcount\msipdfoutput
\ifx\pdfoutput\undefined\else
\ifcase\pdfoutput\else
\ifx\paperwidth\undefined\else
\ifdim\paperheight=0pt\relax\else\pdfpageheight\paperheight\fi
\ifdim\paperwidth=0pt\relax\else\pdfpagewidth\paperwidth\fi
\fi\fi\fi
\begin{document}
\title{Cosmological Constraints on the Generalized Uncertainty Principle from
Redshift-Space Distortions}
\author{Andronikos Paliathanasis}
\email{anpaliat@phys.uoa.gr}
\affiliation{Institute of Systems Science, Durban University of Technology, Durban 4000,
South Africa}
\affiliation{Centre for Space Research, North-West University, Potchefstroom 2520, South Africa}
\affiliation{Departamento de Matem\`{a}ticas, Universidad Cat\`{o}lica del Norte, Avda.
Angamos 0610, Casilla 1280 Antofagasta, Chile}
\affiliation{National Institute for Theoretical and Computational Sciences (NITheCS), South Africa}

\begin{abstract}
We investigate the imprints of the Generalized Uncertainty Principle on cosmological scales by using redshift-space distortion measurements in combination with background cosmological data to determine constraints on the deformation parameter $\beta$. We consider the modified Poisson bracket related to the existence of a minimal length, which leads to a modified Raychaudhuri equation for the standard $\Lambda$CDM model and gives rise to a phenomenological one-parameter dynamical dark energy scenario. Through this modification, we can reveal the effects of the minimal length on the late-time structure of the universe. We employ the $f$ and $f\sigma_8$ measurements of the growth rate combined with background data, including cosmic chronometers, baryon acoustic oscillations and Type Ia supernova observations. The observational constraints reveal a systematically negative value for the deformation parameter $\beta$, with the $\Lambda$CDM limit lying within the 95\% credible interval. When supernova data are included, the Akaike Information Criterion indicates weak-to-strong support in favour of the GUP-modified model depending on the SNIa catalogue, while the Bayesian evidence suggests a weak preference.

\end{abstract}
\maketitle

\section{Introduction}

\label{sec1}

Several approaches of quantum gravity indicate the presence of a fundamental
minimal length on the Planck scale. Such a feature suggests the existence of
an upper bound on the energy scale that can exist in nature which motivates a
modification of the Heisenberg's Uncertainty Principle leading to the
Generalized Uncertainty Principle relation\ (GUP) \cite{Tawfik:2015rva}.
Within the framework of String theory it was found
\cite{Veneziano:1986,Amati:1989} that there exists a shorter measurable
distance following the scattering process at Planck energies. This property
revealed the modification of the Uncertainty Principle with the introduction
of a correction term corresponding to the minimal length \cite{Konishi:1989wk}%
.\ A consistent result was obtained from Black Hole Physics and Doubly Special
Relativity
\cite{Maggiore:1993rv,Maggiore:1993zu,Scardigli:1999jh,Ali:2009zq,Das:2010zf,Chung:2018btu}%
. 

The existence of the minimal length introduces quantum correction terms in
quantum mechanical systems \cite{kempH,vag1}, which open the discussion to
test Planck-scale physics with low-energy experiments
\cite{Pikovski:2011zk,Marin:2013pga}. Corrections on the hydrogen atom
spectrum due to the GUP were derived in \cite{Brau:1999uv}, while the
effects of the GUP on the Harmonic Oscillator were investigated in
\cite{Bosso:2017ndq}, and for the Kepler problem in \cite{Bosso:2018uus}%
.\ Furthermore, applications of the GUP to Newtonian central force systems
were derived in \cite{Scardigli:2016pjs}. For more details we refer to the most
recent discussion in \cite{Casadio:2020rsj}. \ 

The effects of a fundamental minimal length in gravitational physics have been
extensively studied in the literature. In \cite{Adler:2001vs} it was shown that within the
GUP framework there exists a modified mass-temperature profile for the
Hawking radiation which becomes zero for finite mass and prevents the total
evaporation and collapse; a similar result was obtained in
\cite{Banerjee:2010sd}.  Black hole physics can be used to employ constraints
on the GUP deformation parameter \cite{Vagenas:2017vsw}, based on the
algorithm established in \cite{Scardigli:2016pjs}. A similar approach
considered in \cite{Chen:2024mlr,Ali:2024ssf,Fragomeno:2026nnc} where the
deformation parameter was constrained using observable data from the event
horizon telescope. For other studies on the application of the GUP on the
black hole physics we refer the reader to
\cite{Bera:2021lgw,Vagenas:2019rai,Okcu:2019yen,Deb:2016psq,Gregoris:2026hwp,Alencar:2023wyf,Lin:2022doa,Scardigli:2014qka,Heidari:2023ssx,Buoninfante:2020cqz}
and references therein. 

Beyond black hole physics, the implications of GUP have been also investigated
in cosmology and the large structure of the universe. The quantum corrections
related to the minimal length, affects all aspects of the cosmic
history. The effects of GUP in inflation explored in
\cite{Palma:2008tx,Tawfik:2012he,Heydarzadeh:2022lit,Tawfik:2014dza,Maziashvili:2011qs,Paliathanasis:2015cza}%
, while recently in \cite{Heidarian:2025drk}, the observational data from the
Atacama Cosmology Telescope considered to establish bounds on the GUP
deformation parameter. Quantum correction terms can appear in the Raychaudhuri
equation \cite{Vagenas:2017fwa} such that to introduce a cosmological constant
term \cite{Vagenas:2019wzd,Miao:2013wua}, or to modify the cosmological field
equations such that to provide a phenomenological description for the dynamical
character of dark energy
\cite{Kim:2008hz,Ali:2013qza,Okcu:2020ybv,Okcu:2025sby,Luciano:2025ezl,Parsamehr:2024rbi,Paliathanasis:2021egx,Bina:2007wj,Lopez-Aguayo:2023txz,Kouwn:2018rmp,Giardino:2020myz}%
. The modified dynamics of the anisotropic Mixmaster universe within the
framework of GUP investigated in \cite{Segreto:2024gcg}. Furthermore, in
\cite{Paliathanasis:2025mfj}, based on the implication of GUP in the classical
limit as established in \cite{kempH}, it was found that for the inhomogeneous
and anisotropic Szekeres spacetimes, the presence of the minimal length
modifies the classical field equations such that the quantum corrections to
introduce nonlinear dynamical terms in the field equations which can describe
the cosmic acceleration. In the same spirit, in  \cite{Paliathanasis:2025dcr}
it was found that the quadratic GUP can modify the background dynamics for the
$\Lambda$CDM such that to introduce a phenomenological effectively dynamical
dark energy model which can explain the recent cosmological data for the
background. A more general framework for the GUP theory examined later in
\cite{Paliathanasis:2025kmg}. Furthermore, constraints on GUP deformation
parameter using Big Bang nucleosynthesis can be found in
\cite{Luciano:2021vkl,Luo:2024vdd}.

The dynamical nature of dark energy became particularly significant \cite{Arora:2025msq,Kessler:2025kju,Zhang:2025bmk,Silva:2025twg,Silva:2025bnn,Sabogal:2025qhz,Paliathanasis:2025cuc,Nagpal:2025omq,Ormondroyd:2025iaf,You:2025uon,Scherer:2025esj,Yang:2025uyv,Bhattacharjee:2025xeb,Santos:2025wiv,Ibarra-Uriondo:2026zbp,Aboubrahim:2026tks,Li:2026xaz,Li:2025vuh,Wang:2026sqy,Montefalcone:2026iga,Gokcen:2026pkq,Figueruelo:2026eis,Rezaei:2025vhb,Chaudhary:2025uzr,Li:2025ops,Luciano:2025elo} following
the release of the Baryon Acoustic Oscillations (BAO) from the Dark Energy
Spectroscopic Instrument (DESI) DR1 data \cite{DESI:2024uvr,DESI:2024lzq,DESI:2024mwx,DESI:2025zpo,DESI:2025zgx}, and even more so after the
publication of the second release last year \cite{DESI:2025zgx,DESI:2025zpo}. The analysis presented in
\cite{Paliathanasis:2025mfj,Paliathanasis:2025dcr} indicated that the BAO
measurements from DESI DR2 can be used to constrain the GUP deformation
parameter, while background observations show a slight preference for the
GUP-modified $\Lambda$CDM model. In the present study, we employ matter
perturbation data to investigate the imprints of quantum corrections and to
explore the effects of the minimal length on the growth of large-scale structure. For
this analysis, we use redshift-space distortion measurements that probe the growth
of large-scale structure from galaxy redshift surveys in combination with
Supernova observations and cosmic chronometer data. Such a study will, for the
first time, provide constraints on the GUP deformation parameter derived from
structure growth observations. The structure of the paper is as follows.

In Section~\ref{sec2} we review the basic elements of the GUP and show how the
deformed algebra affects the dynamics of classical Hamiltonian systems. This
approach allows the introduction of quantum correction terms in the
Raychaudhuri equation for a class of cosmological models. We review previous
results on FLRW cosmology with a cosmological constant and present how the GUP
modifies the Hubble function in the $\Lambda$CDM model. The GUP-corrected
Hubble function is then tested against cosmological observations, specifically
redshift-space distortion (RSD) measurements. In Section~\ref{sec4} we
describe the datasets used in this work: the RSD data on the growth rate of
matter perturbations, comprising measurements of $f$ and $f\sigma_{8}$ the
Hubble parameter measurements from cosmic chronometers, the BAO data from DESI
DR2, and Type Ia supernova observations from the PantheonPlus, Union3.0, and
DES-Dovekie catalogues. These datasets are analysed both individually and in
combination. We also describe the statistical framework considered for the
parameter estimation and model comparison, such as the Akaike Information
Criterion and the Bayesian evidence to compare the GUP-modified models against
the $\Lambda$CDM.

Section~\ref{sec4a} presents the main results of this study. We report the
observational constraints obtained for the quadratic GUP-modified cosmological
model. In Section~\ref{sec4b} we consider a further modified cosmological
model derived from an extended power-law GUP theory. Both models are compared
with $\Lambda$CDM. Finally, Section~\ref{sec5} summarizes our findings and
draws conclusions.

\section{GUP-modified Cosmology}

\label{sec2}

Consider the existence of a minimal length which leads to the modification of
the Heisenberg's Uncertainty Principle as follows%
\begin{equation}
\Delta X\Delta P\geq\frac{1}{2}\left(  1+\beta\Delta P^{2}\right)
,\label{GUP}%
\end{equation}
in which $\Delta X,~\Delta P$ are the position and momentum uncertainties and
$\beta$ is a suitable parameter with dimension $1/P^{2}$, known as the
deformation parameter. The inequality (\ref{GUP}) is known as the quadratic
GUP \cite{kempH,Vagenas:2019wzd,Vagenas:2020bys}. Parameter $\beta$
is expressed as $\beta={\beta}_{0}/{M_{Pl}^{2}c^{2}=\beta}_{0}\ell_{Pl}%
^{2}/2\hbar^{2}$, $M_{Pl}$ is the Planck mass, $\ell_{Pl}$ ($\approx
10^{-35}~m)$ is the Planck length, $M_{Pl}c^{2}$ ($\approx1.2\times
10^{19}~GeV)$ is the Planck energy and $\beta_{0}$ is a dimensionless coupling
parameter, often assumed to be of order unity in string
theory~\cite{Amati:1989}.

Indeed, in the limit where $\beta_{0}\rightarrow0$, Heisenberg's Uncertainty
Principle is recovered. Although, $\beta_{0}$ is usually considered to have a
positive value, nevertheless, there are various approaches in the literature
\cite{Ong:2018zqn,Ong:2018nzk,Carr:2015nqa,Du:2022mvr,Buoninfante:2019fwr} which investigate the case of negative
value for parameter $\beta_{0}$.

From expression (\ref{GUP}) the following deformation relation for the
canonical algebra can be defined {\cite{Maggiore:1993rv}
\begin{equation}
\lbrack X_{i},P_{j}]=i\left[  \delta_{ij}(1+\beta P^{2})+2\beta P_{i}%
P_{j}\right]  , \label{gup2}%
\end{equation}
while the spatial coordinates commute, as do the momentum components, i.e.
$\left[  X_{i},X_{j}\right]  =0$ and $\left[  P_{i,}P_{j}\right]  =0$.
Nevertheless, in the coordinate representation, for the momentum and the
position, we adopt $X_{i}=x_{i}$ and modify the momentum operator
as~\cite{kempH}
\begin{equation}
P_{i}=(1+\beta p^{2})p_{i}, \label{gup3}%
\end{equation}
where $x_{i},~p_{i}$ denote the standard operators satisfying the canonical
algebra $[x_{i},p_{j}]=i\hbar\delta_{ij}$\thinspace$,$ and $p^{2}=p_{i}p^{i}$. \ 

There are two approaches in the literature for employing the minimal length in
the classical limit, and introduce the quantum effects and corrections.

Consider the classical autonomous Hamiltonian system $\mathcal{H}%
=\mathcal{H}(x,p)$, with Hamiltonian equations \cite{kempH}%
\begin{equation}
\dot{x}_{i}=\left\{  x_{i},\mathcal{H}\right\}  ,~\dot{p}_{i}=\left\{
p_{i},\mathcal{H}\right\}  ,
\end{equation}
then due to the deformation algebra (\ref{gup2}) this leads to modified
Hamilton's equations \cite{vag1}%
\begin{equation}
\dot{x}^{i}=\left(  1-\beta p^{2}\right)  \frac{\partial\mathcal{H}}{\partial
p_{i}},\qquad\dot{p}^{i}=\left(  1-\beta p^{2}\right)  \frac{\partial
\mathcal{H}}{\partial x_{i}},
\end{equation}
via the modification of the Poisson bracket.

In the second approach, the Poisson bracket remains unchanged, and the
Hamiltonian function is considered to be defined $\mathcal{H}=\mathcal{H}%
(x,P)$ from where a quantum correction term is introduced from the
transformation (\ref{gup3}), and in the coordinate representation it follows
\begin{equation}
\mathcal{H}(x,P)=\mathcal{H}_{0}(x,p)+\beta\mathcal{H}_{1}(x,p).
\end{equation}
The Legendre transformation within the framework of the modified Poisson
bracket explored in details in \cite{Bosso:2018uus}. For more details we refer to the discussion in \cite{Bosso:2023aht}.

These two approaches introduce quantum corrections in different ways;
nevertheless, they lead to inequivalent physical systems, which coincide only
in the limit $\beta\rightarrow0$.

\subsection{GUP-modified FLRW Cosmology}

In this work we adopt the first modified Poisson formalism for the study of
the quantum corrections effects driven by the existence of the minimal length
in FLRW cosmology. Following the analysis presented in \cite{Paliathanasis:2025mfj}, for the
Szekeres system with or without the cosmological constant, the modification of
the Poisson bracket leads to quantum correction term within the Raychaudhuri
equation driven by the deformation function $F\left(  p\right)  $. In the
homogeneous and isotropic limit, the Szekeres spacetime reveals the FLRW
cosmological background, and the Friedmann's equations are recovered.

That property was considered in \cite{Paliathanasis:2025dcr} to explore the effects of the
quadratic-inspired GUP within the framework of the standard $\Lambda$CDM
cosmology. Specifically, it has been shown that the cosmological solution for
the $\Lambda$CDM model is modified such that the Hubble function reads
\begin{widetext}
\begin{equation}
H^{2}\left(  z\right)  =H_{0}^{2}\frac{3\Omega_{m0}\left(  1+z\right)  ^{3}%
}{\left(  \frac{\sqrt{1+\frac{16}{9}\beta\left(  1+\beta\right)  }\Omega
_{m0}\left(  1+z\right)  ^{3}}{\Omega_{m0}\left(  1+z\right)  ^{3}+\left(
1-\Omega_{m0}\right)  \left(  1+z\right)  ^{3\left(  1-\sqrt{1+\frac{16}%
{9}\beta\left(  1+\beta\right)  }\right)  }}+\frac{1}{2}\left(  1-\sqrt
{1+\frac{16}{9}\beta\left(  1+\beta\right)  }\right)  -\frac{2}{3}%
\beta\right)  } \label{fe.29a}%
\end{equation}
\end{widetext}
and expanding around $\beta\rightarrow0$, it follows%
\begin{equation}
H^{2}\left(  a\right)  =H_{\Lambda CDM}^{2}\left(  a\right)  +\beta
H_{cor}^{B}\left(  a\right)  +O\left(  \beta^{2}\right)  , \label{fe.30}%
\end{equation}
where $H_{\Lambda CDM}^{2}\left(  a\right)  $ is the Hubble function for the
$\Lambda$CDM, that is,
\begin{equation}
H_{\Lambda CDM}^{2}\left(  a\right)  =H_{0}^{2}\left(  \left(  1-\Omega
_{m0}\right)  +\Omega_{m0}\left(  1+z\right)  ^{3}\right)  \text{,}%
\end{equation}
and $H_{cor}^{B}\left(  a\right)  $ is the quantum correction term which
follows from the minimal length and is given by the expression%
\begin{equation}
\begin{split}
\frac{3H_{\rm cor}^{B}(a)}{2H_0^2} = &\; \Omega_{m0} (1+z)^3 
+ 6 (1-\Omega_{m0}) \\
& + \frac{5 (1-\Omega_{m0})^2}{\Omega_{m0}} (1+z)^{-3} \\
& - 12 (1-\Omega_{m0}) \ln(1+z) \,.
\end{split}
\label{fe.31}
\end{equation}
As discussed in detail in \cite{Paliathanasis:2025dcr}, the corrected Hubble function can be recast in the form of an effective one-parameter dynamical dark energy model, with a phantom-like behavior for negative values of the parameter $\beta$.

In the more general setting in \cite{Paliathanasis:2025kmg} which introduced a power-law GUP
framework, such that the energy density of the matter source is given by
the differential equation%
\begin{equation}
\begin{split}
-(1+z) \frac{d\Omega_{m}}{dz} = &\; 3 \Omega_{m} (\Omega_{m}-1) \\
& + 2 (3 \Omega_{m})^{1-\mu} \Big( (1+\mu)\Omega_{m} - \tfrac{2}{3} - \mu \Big) \beta \,.
\end{split}
\label{fe.31a}
\end{equation}
and the Hubble function
\begin{equation}
\left(  \frac{H\left(  z\right)  }{H_{0}}\right)  ^{2}=\frac{\Omega
_{m0}\left(  1+z\right)  ^{3}}{\Omega_{m}\left(  z\right)  },\,\quad\Omega
_{m}\left(  0\right)  \equiv\Omega_{m0}\,, \label{fe.22}%
\end{equation}
Parameter $\mu$ describes the power of the deformation function $F\left(
p\right)  $, where for $\mu=2,$ the quadratic theory is recovered. We easily
observe this from (\ref{fe.31a}) and (\ref{fe.22}), such that for $\mu=2$ the
analytic expression (\ref{fe.29a}) is recovered.

These GUP-modified cosmological models have previously been investigated to
determine whether they can account for recent background cosmological
observations. In particular, it was examined whether the existence of a
minimal length and the associated deformation of the Poisson algebra can
explain the dynamical behavior of dark energy, and whether cosmological data
can place constraints on the deformation parameter $\beta$. The analysis
indicated weak evidence in favor of the GUP-modified cosmological model, with
the deformation parameter $\beta$ exhibiting a negative median value.

In the present study, we extend this analysis by investigating the impact of
the modified background dynamics on the evolution of matter density
perturbations, with the aim of constraining the deformation parameter $\beta$
using redshift-space distortion measurements, as well as to examine whether the
deviation from the $\Lambda$CDM can explain the cosmological data related to
the matter perturbations.

In this framework, quantum corrections affect the matter perturbation
equations only indirectly, through modifications to the background Hubble
function. Since, at the first order of $\beta$, the dominant effect comes through the Hubble function. Corrections in the perturbations will be of the powe $O(\beta^2)$  or $O(\beta \delta\rho_m)$ Consequently, the evolution of matter perturbations is described by
the differential equation
\begin{equation}
\ddot{\delta}_{m}+2H\dot{\delta}_{m}-\frac{3}{2}H^{2}\Omega_{m}\delta_{m}=0.
\end{equation}

\section{Observational Data}

\label{sec4}

In the following lines we present the observational data which we employed for
this study, as well as the methodology and the priors considered in this work.

We examined the effects of the minimal length in the evolution of the
late-universe, by using different combinations of four different cosmological
measurements as presented below.

\begin{itemize}
\item RSD: We consider redshift-space distortion (RSD) measurements that probe
the growth of large-scale structure from galaxy redshift surveys, which allow
us to constrain the matter perturbations and to test the consistency of
the cosmological model with the observed large-scale structure. Specifically,
these data constrain the quantities $f\left(  z\right)  $ and $f\sigma
_{8}\left(  z\right)  $, where $f\left(  z\right)  $ denotes the growth rate
of matter perturbations defined as $f\left(  z\right)  =\frac{d\ln\delta_{m}%
}{d\ln a}$, $\sigma_{8}\left(  z\right)  $ represents the root-mean-square
amplitude of matter fluctuations in spheres with radius $8h^{-1}Mpc$, that is,
$\sigma_{8}\left(  z\right)  =\sigma_{8,0}\frac{\delta_{m}\left(  z\right)
}{\delta_{m}\left(  0\right)  }$ and $\delta_{m}$ describes the evolution of
the matter perturbations, that is, $\delta_{m}=\frac{\delta\rho_{m}}{\rho_{m}%
}$. In this work we employ two RSD datasets. First, we consider eleven direct
measurements of the growth rate$~f\left(  z\right)  $ at different redshifts
within the range $0.013\leq z\leq1.40$, obtained from multiple galaxy surveys,
as presented in Table 1 of \cite{Avila:2022xad,Escobal:2026lnp}. Furthermore, we consider the twenty
measurements of the $f\sigma_{8}\left(  z\right)  $ as they are summarized in
Table 2 of \cite{Avila:2022xad} for redshifts in the range $0.02\leq z\leq 1.944$.

\item CC: The Cosmic Chronometers\ (CC) are used as direct model independent
measurements of the Hubble parameters. The latter is determined by the age
difference $\frac{dz}{dt}$ between galaxies at neighboring redshifts, where
the Hubble parameter is given by the formula $H\left(  z\right)  =-\frac
{1}{\left(  1+z\right)  }\frac{dz}{dt}$. In this study we consider the 31
measurements of $H\left(  z\right)  $ within the redshifts $0.09\leq
z\leq1.965$, as followed from the analysis presented in \cite{moresco2020setting}. The
likelihood analysis incorporates the full covariance
matrix\footnote{https://gitlab.com/mmoresco/CCcovariance} following the
approach described in \cite{Moresco:2020fbm}.

\item BAO: We employ the latest catalogue for the BAO measurements as given by
the Dark Energy Spectroscopic Instrument Data Release 2 (DESI DR2)
\cite{DESI:2025zgx,DESI:2025zpo}. This catalogue provides
observable distance measurements at seven different redshifts, normalized by
the sound horizon at the baryon drag epoch $r_{drag}$. The observables are
the transverse comoving angular distance ratio, $D_{M}r_{drag}^{-1},~$the
volume-averaged distance ratio, $D_{V}r_{drag}^{-1}$ and and the Hubble
distance ratio $D_{H}r_{drag}^{-1}.$

\item SNIa: We use three different supernova Type Ia compilations,
PantheonPlus (PP) \cite{Brout:2022vxf}, Union3.0 (U3) \cite{rubin2023union}
and DES-Dovekie (DESD) \cite{DES:2025sig}. The PP sample contains 1550 SNIa
observations in the redshift range $10^{-3}<z<2.27$, while the U3 includes
2087 SNIa covering a comparable redshift range. Finally, the DESD catalogue is
based on a reanalysis of the five-year Dark Energy Survey supernova program
(DES-SN5YR), and includes 1820 SNIa events in the redshift range$~z<1.13$.
\end{itemize}

In order to perform the constraints, we employ the Bayesian inference framework
Cobaya\footnote{https://cobaya.readthedocs.io/} \cite{cob1,cob2}, with a
custom theory for the derivation of the observables, and the PolyChord nested
sampling algorithm \cite{poly1,poly2}. For any combination of the above
dataset we define the maximum value of the $\mathcal{L}_{\max}$ given by the
expression
\[
\mathcal{L}_{\max}^{total}=\mathcal{L}_{\max}^{Data~A}\times\mathcal{L}_{\max
}^{Data~B}\times...,
\]
where $\mathcal{L}_{\max}=\exp\left(  -\frac{1}{2}\chi_{\min}^{2}\right)  $.

The free parameters for the $\Lambda$CDM are the $\left\{  H_{0},\Omega
_{m0},r_{drag},\sigma_{8,0}\right\}  $, while the GUP-modified model
introduces the additional parameter $\beta$. For the $H_{0}$ we considered the
prior, $H_{0}\in\left[  60,80\right]  $, for the $\Omega_{m0}\in\left[
0.1,0.5\right]  $, the $r_{drag}$ $\in\left[  120,170\right]  $, the
$\sigma_{8,0}\in\left[  0.3,1.2\right]  $ and for deformation parameter
$\beta\in\left[  -0.3,0.2\right]  $. Moreover,\ for the analysis of numerical
outcomes we make use of the GetDist
library\footnote{https://getdist.readthedocs.io/}~\cite{getd}.

Finally, in order to investigate whether the GUP-modified standard cosmology
provides a better fit to the data than the $\Lambda$CDM, we employ two
statistical criteria designed for the comparison of models with a different
number of free parameters. In particular, we use the Akaike Information
Criterion (AIC) \cite{AIC} and the Bayesian evidence \cite{AIC2}. Both
criteria penalize models with a larger number of free parameters, thereby
preventing the preference of more complex models unless they provide a
significantly improved fit to the data.

For each dataset and for each model, we calculate the $AIC$ parameter from the
formula $AIC=\chi_{\min}^{2}+2\mathcal{N}$, where $\mathcal{N}$ is the
dimension of the parametric space. For the $\Lambda$CDM, we have
$\mathcal{N}_{\Lambda}=4$ and for the GUP-modified model $\mathcal{N}_{GUP}%
=5$. Thus, from the difference of the $AIC$ parameters
\[
\Delta AIC=AIC_{GUP}-AIC_{\Lambda},
\]
or equivalently%
\[
\Delta AIC=\chi_{\min}^{2~GUP}-\chi_{\min}^{2~\Lambda}+2,
\]
and the use of the Akaike's scale, we conclude that if $\left\vert \Delta
AIC\right\vert <2$ the two models are statistically indistinguishable.
Moreover, if $-4<\Delta AIC<-2$, the data provide weak support for the
GUP-modified model, while for $2<\Delta AIC<4$, the data provide a marginal
evidence in favor of the $\Lambda$CDM. Nevertheless, if$~-7$ $<\Delta AIC<-4$,
there is a strong support in favor of the GUP-modified model, while for
$4<\Delta AIC<7$, the strong support is provided to the $\Lambda$CDM, while
for $\left\vert \Delta AIC\right\vert >7$ there is a clear evidence in favor
to the model with lower $AIC$.

The Bayesian evidence $\ln Z~$represents the marginal likelihood of the data
given a model and it is obtained by integrating the likelihood function over
the full parameter space weighted by the prior distributions. The Bayesian
evidence can be used for the model comparison with the consideration of
Jeffrey's scale \cite{AIC2} and the difference $\Delta\ln Z=\ln Z_{GUP}-\ln
Z_{\Lambda}$. In this analysis the Bayesian evidence is directly provided by
the PolyChord nested sampler. According to Jeffrey's scale, if $\Delta\ln
Z<1$, the two models are comparable, while for $1<\Delta\ln Z<2.5,$ there is a
weak favor for the GUP-modified model, while for $-2.5<\Delta\ln Z<-1$ the
data set provides a weak favor for the $\Lambda$CDM. Furthermore, if
$2.5<\Delta\ln Z<5$, the evidence in favor of the GUP-model is moderate, while
for $-5<\Delta\ln Z<-2.5$, the moderate evidence is for the $\Lambda$CDM.
Finally, for $\Delta\ln Z>5$, the data provide strong support for
the GUP-modified model, nevertheless, the support is for the $\Lambda$CDM when
$\Delta\ln Z<-5$.

\section{Observational Constraints of the Quadratic GUP}

\label{sec4a}

In this Section we present the numerical outcomes obtained from the
observational constraints. We performed the analysis for different combinations
of data, categorized in two families, the datasets without SNIa, and with
SNIa. In Tables \ref{tab1} and \ref{tab2} we present the mean values and the
marginalized posterior credible intervals (CI) at the $68\%$ level as derived
from the MCMC chains for the different combinations of the above presented
data sets for the GUP-modified cosmological model. Furthermore the median
values for the free parameters obtained for the $\Lambda$CDM$\ $are presented.%

\begin{table*}[tbp] \centering
\caption{Numerical outcomes for the quadratic GUP-modified model and comparison of the statistical parameters with respect to the $\Lambda$CDM. Within parenthesis are the mean values for the $\Lambda$CDM (1/2).}%
\begin{tabular}
[c]{cccc}\hline\hline
\textbf{Dataset} & $\mathbf{H}_{0}\mathbf{~}$ & $\mathbf{\Omega}_{m0}$ &
$\mathbf{r}_{drag}$\\\hline
\multicolumn{4}{c}{\textbf{Without SNIa}}\\\hline
${\small f\sigma}_{8}$ & $-$ & $0.296_{-0.073}^{+0.061}\left(  0.300\right)  $
& $-$\\
${\small f+f\sigma}_{8}$ & $-$ & $0.301_{-0.041}^{+0.035}\left(  0.278\right)
$ & $-$\\\hline
{\small CC+BAO} & $68.4_{-2.8}^{+2.8}\left(  69.3\right)  $ & $0.284_{-0.034}%
^{+0.040}\left(  0.297\right)  $ & $147.3_{-3.5}^{+3.5}\left(  146.7\right)
$\\
{\small CC+BAO+}$f\sigma_{8}$ & $68.6_{-2.9}^{+2.6}\left(  69.4\right)  $ &
$0.287_{-0.034}^{+0.034}\left(  0.297\right)  $ & $147.2_{-3.4}^{+3.4}\left(
146.7\right)  $\\
{\small CC+BAO+}$f${\small +}$f\sigma_{8}$ & $68.3_{-2.0}^{+2.0}\left(
69.5\right)  $ & $0.284_{-0.015}^{+0.015}\left(  0.293\right)  $ &
$147.1_{-3.4}^{+3.4}\left(  146.9\right)  $\\\hline
\multicolumn{4}{c}{\textbf{With SNIa}}\\\hline
{\small PP+CC+BAO} & $67.9_{-1.7}^{+1.7}\left(  68.7\right)  $ &
$0.281_{-0.021}^{+0.021}\left(  0.311\right)  $ & $147.2_{-3.5}^{+3.5}\left(
146.5\right)  $\\
{\small PP+CC+BAO+}$f\sigma_{8}$ & $68.0_{-1.7}^{+1.7}\left(  68.8\right)  $ &
$0.281_{-0.022}^{+0.019}\left(  0.310\right)  $ & $147.2_{-3.6}^{+3.3}\left(
146.5\right)  $\\
{\small PP+CC+BAO+}$f${\small +}$f\sigma_{8}$ & $68.0_{-1.7}^{+1.7}\left(
68.9\right)  $ & $0.283_{-0.022}^{+0.020}\left(  0.307\right)  $ &
$147.2_{-3.4}^{+3.4}\left(  146.7\right)  $\\\hline
{\small U3+CC+BAO} & $66.8_{-1.8}^{+1.8}\left(  68.7\right)  $ &
$0.261_{-0.025}^{+0.022}\left(  0.312\right)  $ & $147.3_{-3.5}^{+3.5}\left(
146.5\right)  $\\
{\small U3+CC+BAO+}$f\sigma_{8}$ & $66.9_{-1.8}^{+1.8}\left(  68.7\right)  $ &
$0.263_{-0.018}^{+0.020}\left(  0.311\right)  $ & $147.2_{-3.6}^{+3.3}\left(
146.5\right)  $\\
{\small U3+CC+BAO+}$f${\small +}$f\sigma_{8}$ & $67.2_{-1.7}^{+1.7}\left(
68.9\right)  $ & $0.279_{-0.016}^{+0.016}\left(  0.306\right)  $ &
$147.2_{-3.5}^{+3.5}\left(  146.7\right)  $\\\hline
{\small DD+CC+BAO} & $67.8_{-1.7}^{+1.7}\left(  68.4\right)  $ &
$0.275_{-0.021}^{+0.021}\left(  0.313\right)  $ & $147.2_{-3.5}^{+3.5}\left(
147.0\right)  $\\
{\small DD+CC+BAO+}$f\sigma_{8}$ & $67.8_{-1.7}^{+1.7}\left(  68.4\right)  $ &
$0.276_{-0.021}^{+0.019}\left(  0.313\right)  $ & $147.2_{-3.5}^{+3.5}\left(
146.9\right)  $\\
{\small DD+CC+BAO+}$f${\small +}$f\sigma_{8}$ & $67.8_{-1.6}^{+1.6}\left(
68.6\right)  $ & $0.281_{-0.015}^{+0.015}\left(  0.309\right)  $ &
$147.2_{-3.4}^{+3.4}\left(  147.0\right)  $\\\hline\hline
\end{tabular}
\label{tab1}%
\end{table*}%
%

\begin{table*}[tbp] \centering
\caption{Numerical outcomes for the quadratic GUP-modified model and comparison of the statistical parameters with respect to the $\Lambda$CDM. Within parenthesis are the mean values for the $\Lambda$CDM (2/2).}%
\begin{tabular}
[c]{cccc}\hline\hline
\textbf{Dataset} & $\mathbf{\beta}$ & $\mathbf{\sigma}_{8,0}$ & $\mathbf{S}%
_{8,0}$\\\hline
\multicolumn{4}{c}{\textbf{Without SNIa}}\\\hline
${\small f\sigma}_{8}$ & $-0.040_{-0.12}^{+0.15}$ & $0.837_{-0.180}%
^{+0.091}\left(  0.799\right)  $ & $0.815_{-0.076}^{+0.099}\left(
0.793\right)  $\\
${\small f+f\sigma}_{8}$ & $-0.054_{-0.037}^{+0.073}$ & $0.830_{-0.039}%
^{+0.039}\left(  0.808\right)  $ & $0.830_{-0.06}^{+0.06}\left(  0.776\right)
$\\\hline
{\small CC+BAO} & $-0.018_{-0.054}^{+0.054}$ & $-$ & $-$\\
{\small CC+BAO+}$f\sigma_{8}$ & $-0.014_{-0.052}^{+0.052}$ & $0.821_{-0.085}%
^{+0.053}\left(  0.796\right)  $ & $0.798_{-0.033}^{+0.033}\left(
0.791\right)  $\\
{\small CC+BAO+}$f${\small +}$f\sigma_{8}$ & $-0.021_{-0.023}^{+0.023}$ &
$0.823_{-0.037}^{+0.033}\left(  0.799\right)  $ & $0.800_{-0.028}%
^{+0.028}\left(  0.789\right)  $\\\hline
\multicolumn{4}{c}{\textbf{With SNIa}}\\\hline
{\small PP+CC+BAO} & $-0.028_{-0.017}^{+0.015}$ & $-$ & $-$\\
{\small PP+CC+BAO+}$f\sigma_{8}$ & $-0.028_{-0.017}^{+0.015}$ &
$0.832_{-0.039}^{+0.035}\left(  0.790\right)  $ & $0.803_{-0.025}%
^{+0.025}\left(  0.802\right)  $\\
{\small PP+CC+BAO+}$f${\small +}$f\sigma_{8}$ & $-0.026_{-0.012}^{+0.012}$ &
$0.828_{-0.032}^{+0.032}\left(  0.792\right)  $ & $0.804_{-0.025}%
^{+0.025}\left(  0.800\right)  $\\\hline
{\small U3+CC+BAO} & $-0.057_{-0.023}^{+0.021}$ & $-$ & $-$\\
{\small U3+CC+BAO+}$f\sigma_{8}$ & $-0.055_{-0.022}^{+0.022}$ &
$0.872_{-0.052}^{+0.043}\left(  0.789\right)  $ & $0.814_{-0.026}%
^{+0.026}\left(  0.803\right)  $\\
{\small U3+CC+BAO+}$f${\small +}$f\sigma_{8}$ & $-0.041_{-0.016}^{+0.016}$ &
$0.843_{-0.034}^{+0.034}\left(  0.792\right)  $ & $0.813_{-0.026}%
^{+0.026}\left(  0.799\right)  $\\\hline
{\small DD+CC+BAO} & $-0.033_{-0.017}^{+0.014}$ & $-$ & $-$\\
{\small DD+CC+BAO+}$f\sigma_{8}$ & $-0.033_{-0.016}^{+0.014}$ &
$0.840_{-0.037}^{+0.037}\left(  0.789\right)  $ & $0.804_{-0.025}%
^{+0.025}\left(  0.805\right)  $\\
{\small DD+CC+BAO+}$f${\small +}$f\sigma_{8}$ & $-0.029_{-0.012}^{+0.012}$ &
$0.832_{-0.031}^{+0.031}\left(  0.790\right)  $ & $0.805_{-0.025}%
^{+0.025}\left(  0.802\right)  $\\\hline\hline
\end{tabular}
\label{tab1b}%
\end{table*}%

\subsection{Without SNIa}

We first consider dataset combinations that do not include SNIa. When only the
RSD data are introduced, i.e. $f\sigma_{8}$ and $f+f\sigma_{8}$, the
constraints weak and exhibit large uncertainties. Specifically for the combined
dataset $f+f\sigma_{8}$ we obtained $\Omega_{m0}=0.301_{-0.041}^{+0.035}$, and
$\sigma_{8,0}=0.830_{-0.039}^{+0.039}$,~$S_{8,0}=0.830_{-0.06}^{+0.06}$.
Moreover, the deformation parameter $\beta$ is weakly constrained by the
growth data alone, with the mean value to be negative, that is, \ $\beta
=-0.054_{-0.037}^{+0.073}$.

The introduction of the CC and the BAO, improves the constraints, and reduce
the uncertainties on the posterior variables. For the combined data
{\small CC+BAO+}$f${\small +}$f\sigma_{8}$ we find $H_{0}=68.3_{-2.0}^{+2.0}%
$,~$\Omega_{m0}=0.284_{-0.015}^{+0.015}$,~$\sigma_{8,0}=0.823_{-0.037}%
^{+0.033}$~and $S_{8,0}=0.800_{-0.028}^{+0.028}$, while for the deformation
parameter $\beta$ it follows $\beta=-0.021_{-0.023}^{+0.023}$.
The confidence space of the free parameters for the latter combined data are presented in Fig. \ref{contour0}.

\begin{figure}[h]
\centering\includegraphics[width=0.50\textwidth]{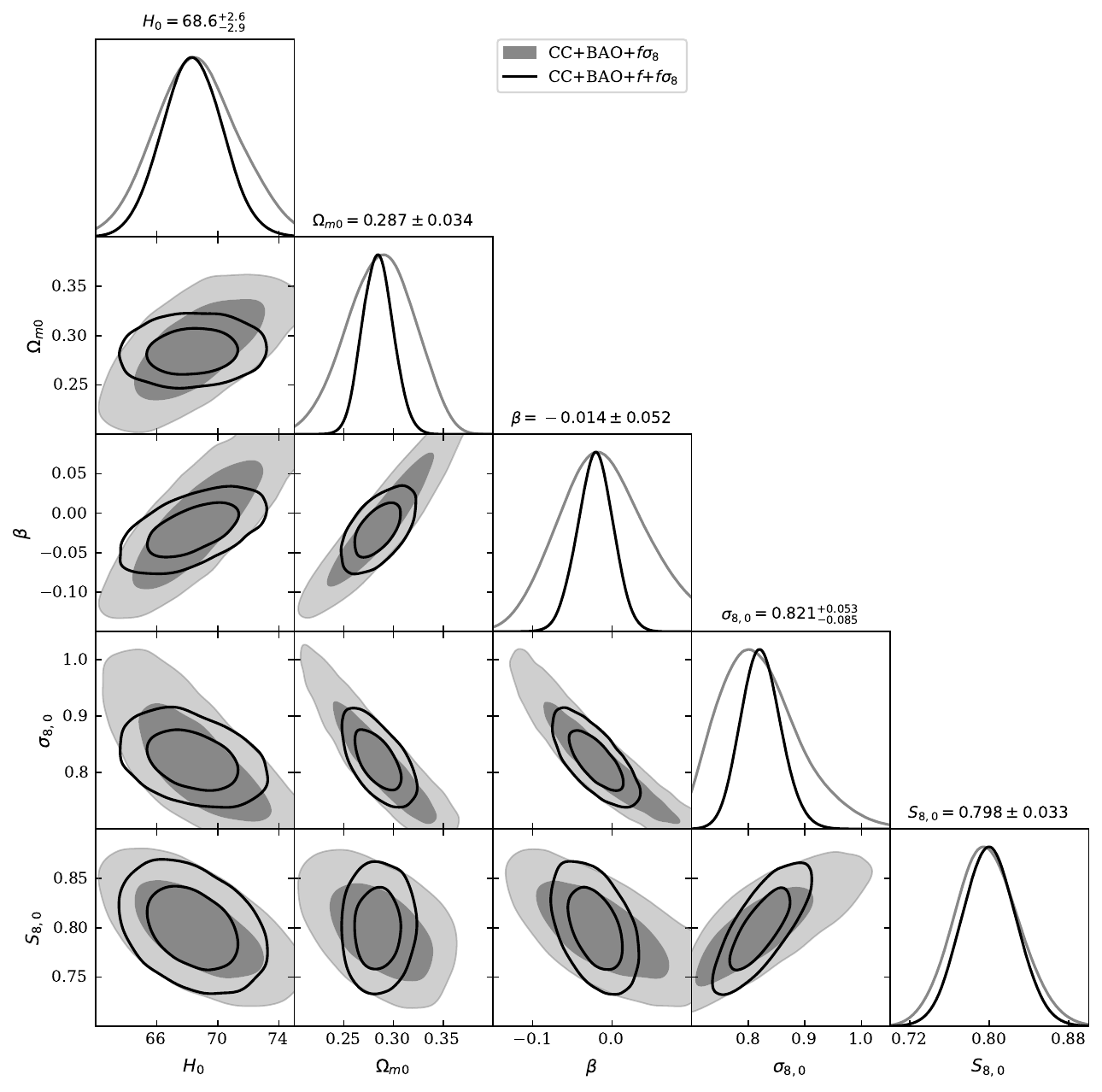}\caption{Confidence
space for the posterior parameters GUP-modified model for the data sets without SNIa.}%
\label{contour0}%
\end{figure}

\subsection{With SNIa}

The introduction of the SNIa data on the combined data, leads to significantly
smaller uncertainties, and tightens the constraints for the free parameters. The
combined data\ indicate negative values for the deformation parameter $\beta$
as presented in Fig. \ref{pla2a}, with the limit of the $\Lambda$CDM, i.e.
$\beta=0$ to be at the $95\%$~CI. \ The small values of $\beta$ confirm the
GUP corrections on the $\Lambda$CDM at the late universe, due to the existence
of minimal length at the quantum level.

\begin{figure}[h]
\centering\includegraphics[width=0.5\textwidth]{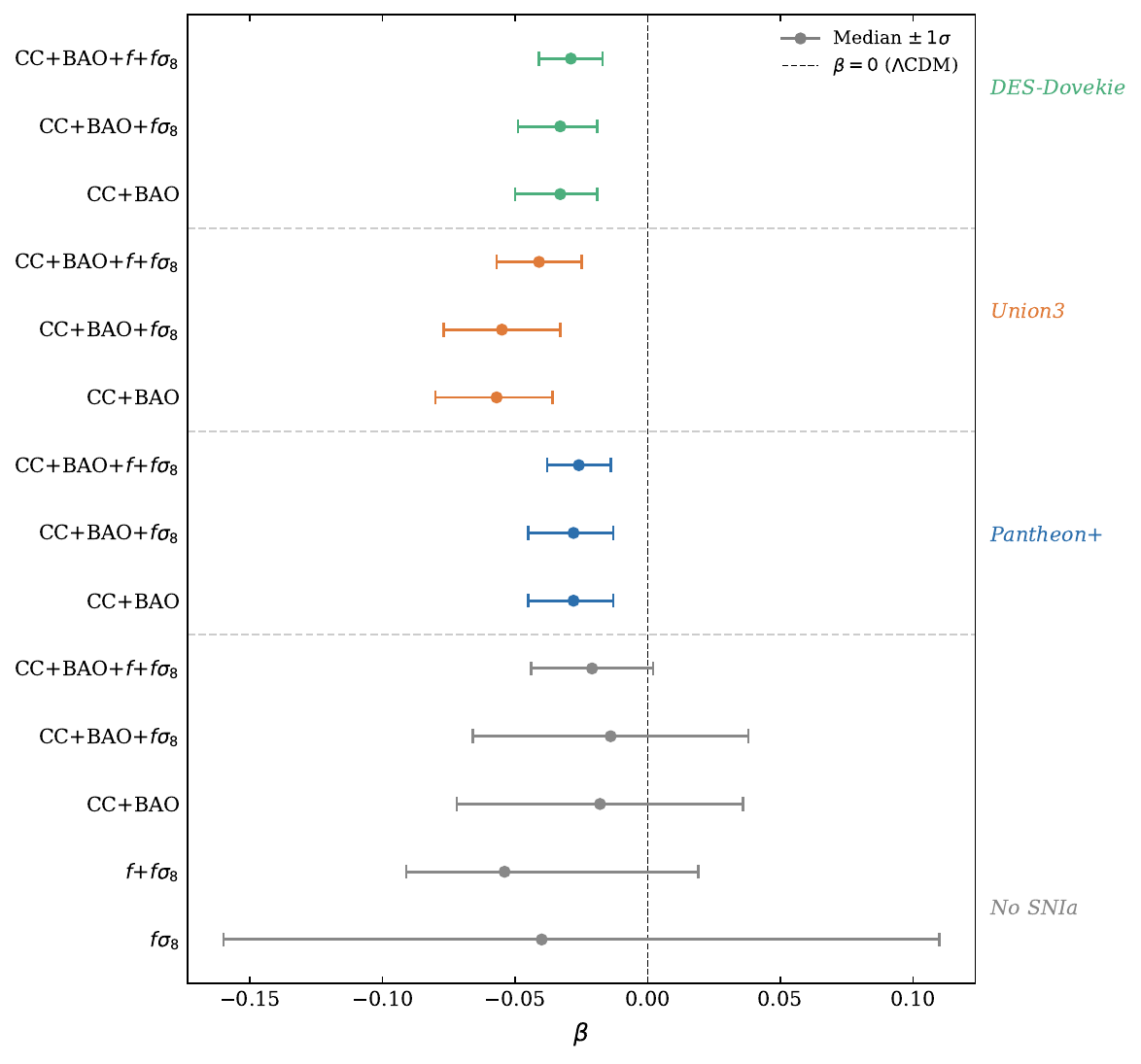}\caption{Marginalized
constraints on the deformation parameter $\beta$ for the different dataset
combinations. We observe that when the supernova data are introduced, there is
a consistent preference for a negative value~of $\beta$ at the $68\%$ confidence
level.}%
\label{pla2a}%
\end{figure}

When the PP SNIa catalogue is introduced, and for the combined data set
{\small PP+CC+BAO+}$f${\small +}$f\sigma_{8}$ we derive the $H_{0}%
=68.0_{-1.7}^{+1.7}$, $~\Omega_{m0}=0.283_{-0.022}^{+0.020}$, $\sigma
_{8,0}=0.828_{-0.032}^{+0.032}$, $S_{8,0}=0.804_{-0.025}^{+0.025}$, and
$\beta=-0.026_{-0.012}^{+0.012}$. On the other hand for the dataset
{\small U3+CC+BAO+}$f${\small +}$f\sigma_{8}$ the analysis of the MCMC
chains reveal $H_{0}=67.2_{-1.7}^{+1.7}$, $~\Omega_{m0}=0.279_{-0.016}%
^{+0.016}$, $\sigma_{8,0}=0.843_{-0.034}^{+0.034},$ $S_{8,0}=0.813_{-0.026}%
^{+0.026}$, and $\beta=-0.041_{-0.016}^{+0.016}$. Finally, the application of
the recent DESD catalogue, and the dataset {\small DD+CC+BAO+}$f$%
{\small +}$f\sigma_{8}$ provides the constraints for the physical parameters
$H_{0}=67.8_{-1.6}^{+1.6}$, $~\Omega_{m0}=0.281_{-0.015}^{+0.015}$,
$\sigma_{8,0}=0.832_{-0.031}^{+0.031},$ $S_{8,0}=0.805_{-0.025}^{+0.025}$, and
$\beta=-0.029_{-0.012}^{+0.012}$.

In Figs. \ref{contour1}, \ref{contour2} and \ref{contour3}  we present the contours for the confidence
space of the free parameters

\begin{figure}[h]
\centering\includegraphics[width=0.5\textwidth]{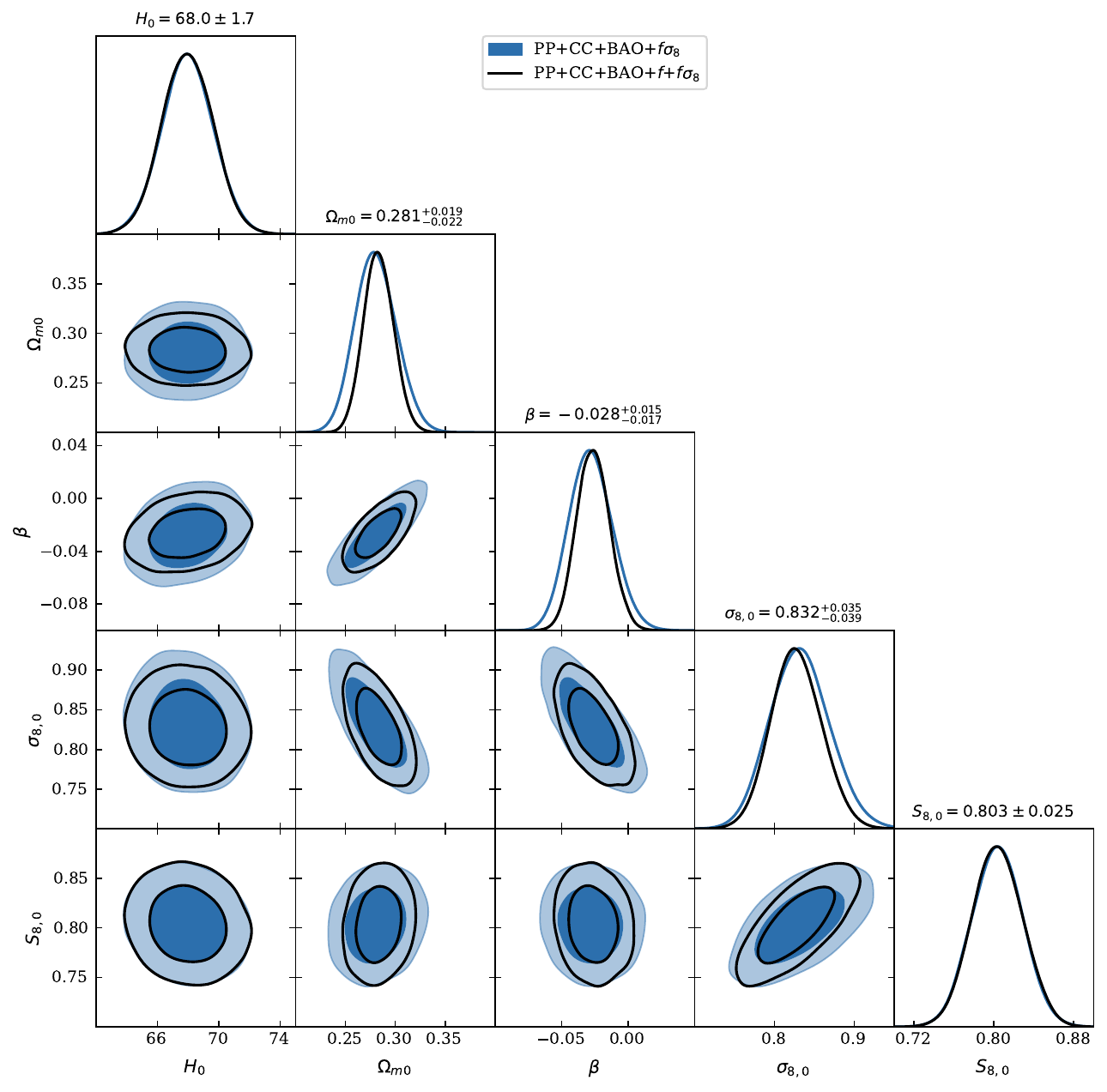}\caption{Confidence
space for the posterior parameters GUP-modified model for the data sets combined with the PP catalogue.}%
\label{contour1}%
\end{figure}

\begin{figure}[h]
\centering\includegraphics[width=0.5\textwidth]{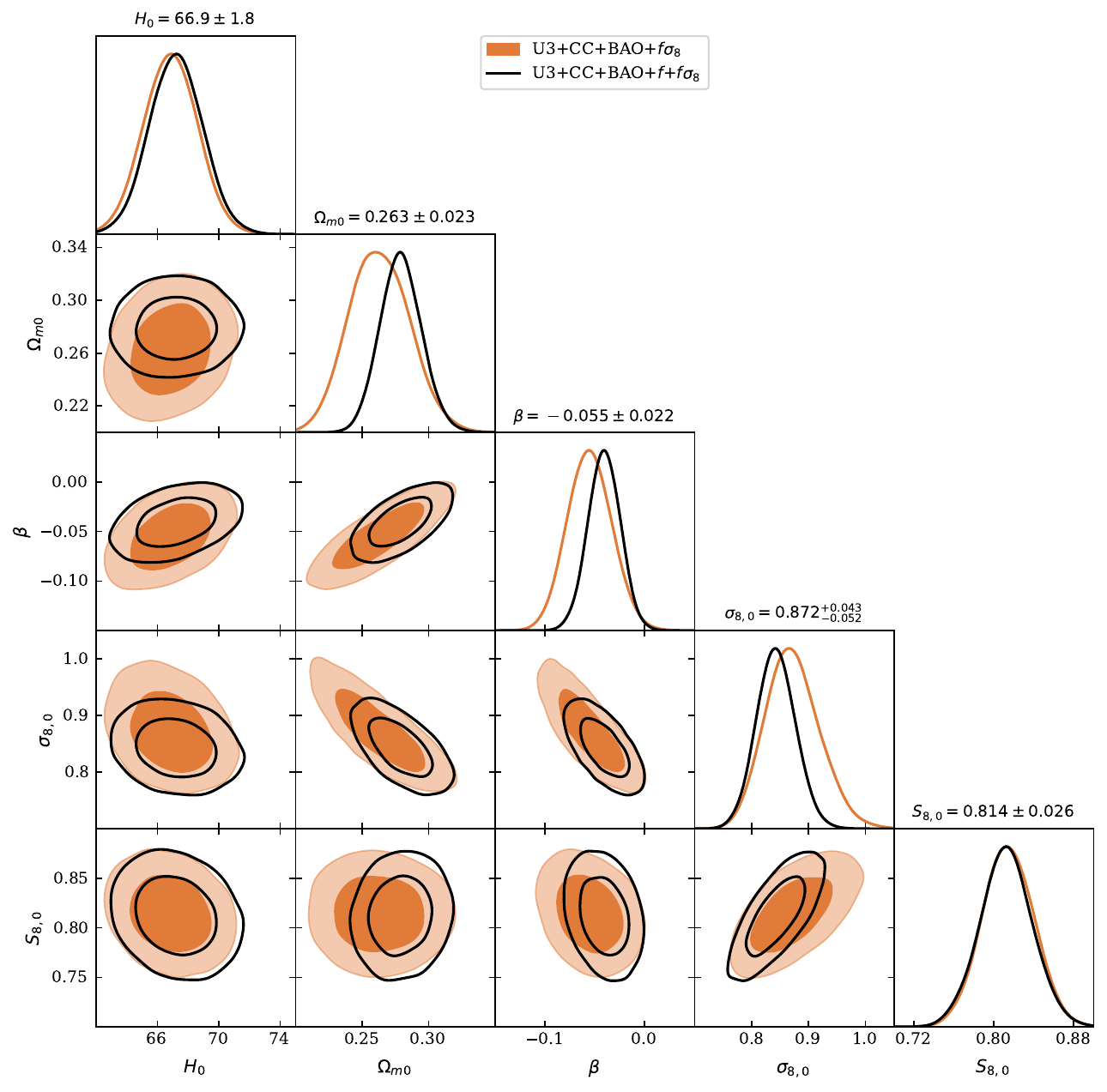}\caption{Confidence
space for the posterior parameters GUP-modified model for the data sets combined with the U3 catalogue.}%
\label{contour2}%
\end{figure}

\begin{figure}[h]
\centering\includegraphics[width=0.5\textwidth]{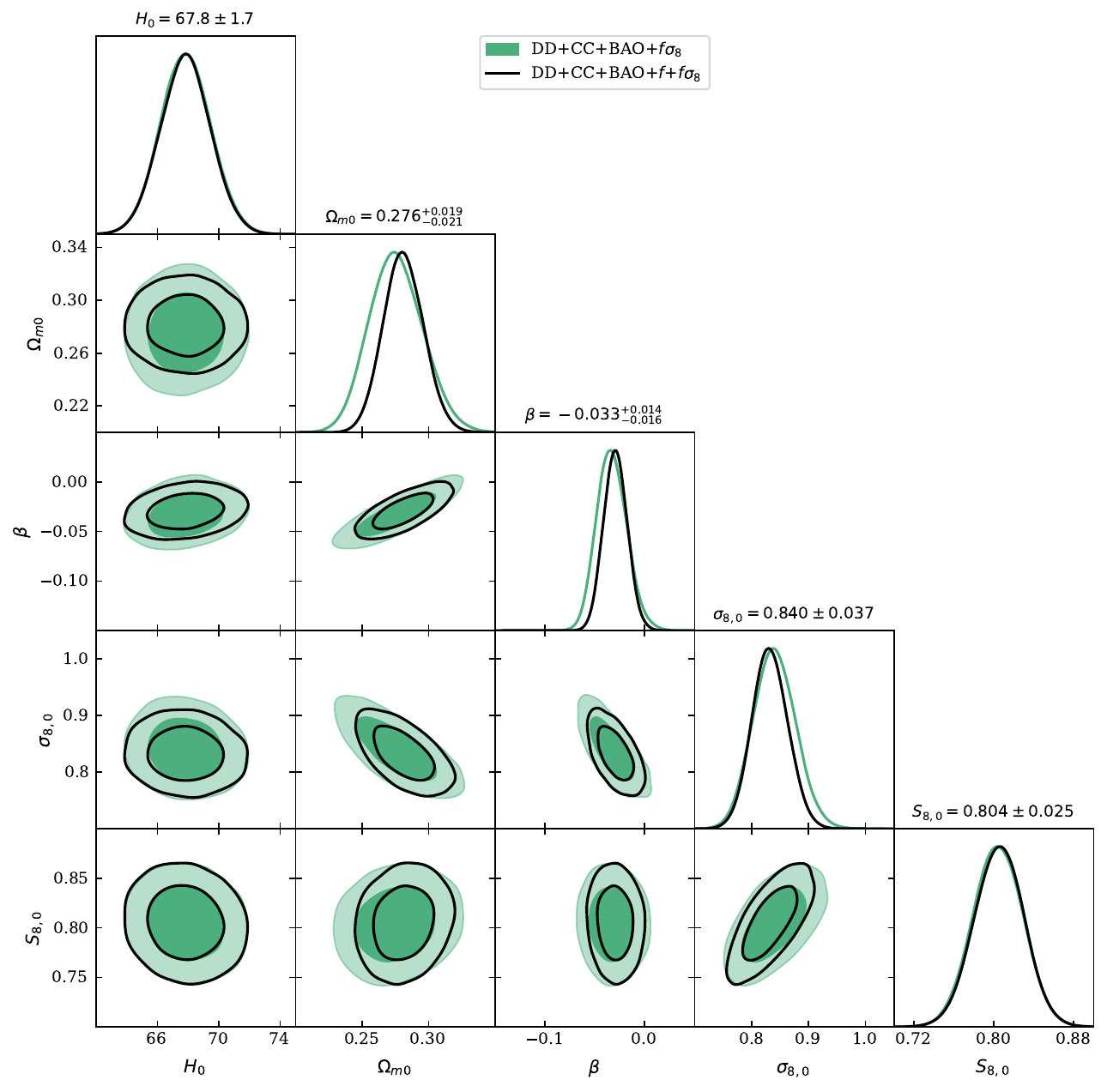}\caption{Confidence
space for the posterior parameters GUP-modified model for the data sets combined with the DD catalogue.}%
\label{contour3}%
\end{figure}

The comparison of the obtained physical parameters for the GUP model, with
that of the $\Lambda$CDM, reveals a systematically smaller value for the $H_{0}$,
with differences of approximately $0.6-1.7$ $kms^{-1}Mpc^{-1}$, which is
closer to the Planck 2018 value for the $H_{0}$. A similar behaviour is
observed for the $\Omega_{m0}$ value, which is smaller in the GUP-modified theory.

\begin{figure*}[h]
\centering\includegraphics[width=1\textwidth]{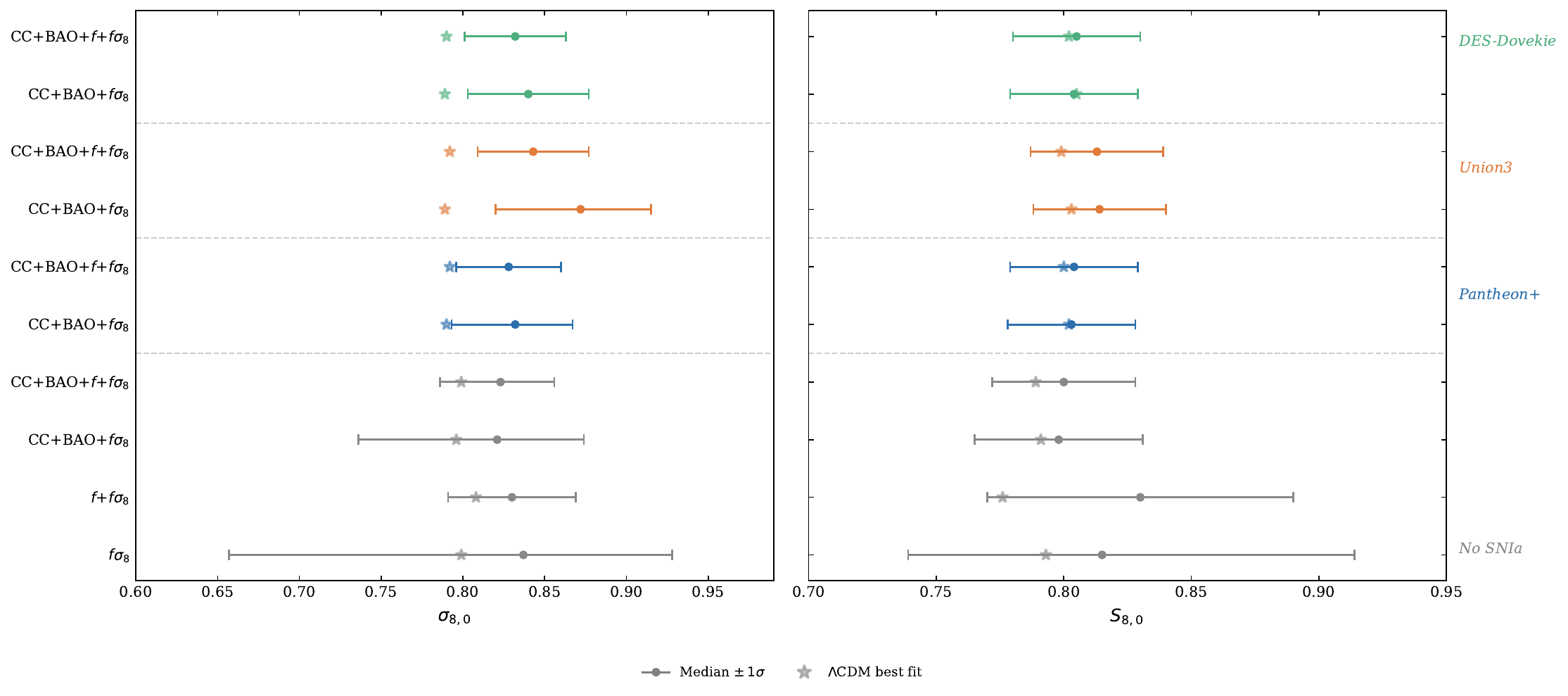}\caption{Marginalized
constraints on the parameters $\sigma_{8,0}$ and $S_{8,0}~$as obtained for the
GUP model for the different dataset combinations. With stars are marked the
mean values for the $\Lambda$CDM. }%
\label{pla3a}%
\end{figure*}

Furthermore, as far as the $\sigma_{8,0}$ and $S_{8,0}$ parameters are
concerned, the obtained values are lower from that of the Planck 2018,
however, the shift is not sufficient to solve the $S_{8,0}$ tension. The
derived values include the mean values given by the $\Lambda$CDM within the
credible intervals at the $68\%$ level for the majority of the data sets as
presented in Fig. \ref{pla3a}.

\subsection{Statistical Comparison}%

\begin{table*}[tbp] \centering
\caption{Statistical comparison of the quadratic GUP-modified model with standard $\Lambda$CDM using Bayesian Evidence and AIC}%
\begin{tabular}
[c]{cccccc}\hline\hline
\textbf{Dataset} & $\Delta\chi_{\min}^{2}$ & $\Delta\left(  AIC\right)  $ &
\textbf{Akaike's scale} & $\Delta\left(  \ln Z\right)  $ & \textbf{Jeffrey's
scale}\\\hline
\multicolumn{6}{c}{\textbf{Without SNIa}}\\\hline
${\small f\sigma}_{8}$ & $-0.08$ & $+1.92$ & Inconclusive & $-0.23$ &
Inconclusive\\
${\small f+f\sigma}_{8}$ & $-0.28$ & $+1.72$ & Inconclusive & $-0.70$ &
Inconclusive\\\hline
{\small CC+BAO} & $-1.29$ & $+0.71$ & Inconclusive & $-0.05$ & Inconclusive\\
{\small CC+BAO+}$f\sigma_{8}$ & $-1.36$ & $+0.64$ & Inconclusive & $+0.28$ &
Inconclusive\\
{\small CC+BAO+}$f${\small +}$f\sigma_{8}$ & $-1.95$ & $+0.05$ & Inconclusive &
$-0.39$ & Inconclusive\\\hline
\multicolumn{6}{c}{\textbf{With SNIa}}\\\hline
{\small PP+CC+BAO} & $-4.10$ & $-2.10$ & Weak favor for GUP & $+0.15$ &
Inconclusive\\
{\small PP+CC+BAO+}$f\sigma_{8}$ & $-4.05$ & $-2.05$ & Weak favor for GUP &
$+0.36$ & Inconclusive\\
{\small PP+CC+BAO+}$f${\small +}$f\sigma_{8}$ & $-5.41$ & $-3.41$ & Weak favor
for GUP & $+1.77$ & Weak favor for GUP\\\hline
{\small U3+CC+BAO} & $-6.97$ & $-4.97$ & Strong favor for GUP & $+2.21$ & Weak
favor for GUP\\
{\small U3+CC+BAO+}$f\sigma_{8}$ & $-6.72$ & $-4.72$ & Strong favor for GUP &
$+2.08$ & Weak favor for GUP\\
{\small U3+CC+BAO+}$f${\small +}$f\sigma_{8}$ & $-7.26$ & $-5.26$ & Strong
favor for GUP & $+1.72$ & Weak favor for GUP\\\hline
{\small DD+CC+BAO} & $-4.29$ & $-2.29$ & Weak favor for GUP & $+0.36$ &
Inconclusive\\
{\small DD+CC+BAO+}$f\sigma_{8}$ & $-4.22$ & $-2.22$ & Weak favor for GUP &
$+1.26$ & Weak favor for GUP\\
{\small DD+CC+BAO+}$f${\small +}$f\sigma_{8}$ & $-5.63$ & $-3.63$ & Weak
favor for GUP & $+2.02$ & Weak favor for GUP\\\hline\hline
\end{tabular}
\label{tab2}%
\end{table*}%

In Table \ref{tab2} we present the $\Delta\chi_{\min}^{2}$, the $\Delta\left(
AIC\right)  $ and the $\Delta\left(  \ln Z\right)  $ between the GUP model and
the $\Lambda$CDM, while in Fig. \ref{pla1}. \ we provide a graphical
representation of the table. We remark that for all the datasets the GUP model
provided a smaller value for the $\chi_{\min}^{2}$.

For datasets that do not include SNIa observations, the values of
$\Delta\left(  AIC\right)  $ and $\Delta\left(  \ln Z\right)  ~$indicate that
the two models provide statistically comparable fits to the data, with no
significant preference for either model. The conclusions are different when
SNIa data are included.

When the PP catalogue is introduced, the Akaike criterion suggests a weak
preference for the GUP model across all dataset combinations. However, the
Bayesian evidence indicates only a weak preference for the GUP model, and only
for the combination of data {\small PP+CC+BAO+}$f${\small +}$f\sigma_{8}$.
Stronger deviations from $\Lambda$CDM are found when the U3 catalogue is
introduced. In this case, the $\Delta\left(  AIC\right)  $ values suggest a
strong preference for the GUP model across all dataset combinations. The
Bayesian evidence, suggests a weak preference for the GUP model for the
different combinations of data.

Finally, a similar behavior is found for the DD catalogue. The Akaike
criterion and the Bayesian evidence suggest a weak preference in favor of GUP when the RSD data are introduced.

Although the $\Delta\chi_{\min}^{2}~$is systematically negative, the inclusion of
SNIa observations, and specifically the use of the U3 and DD catalogues,
indicate a preference for the GUP model, with the Bayesian evidence
providing weak support.

\begin{figure*}[h]
\centering\includegraphics[width=1\textwidth]{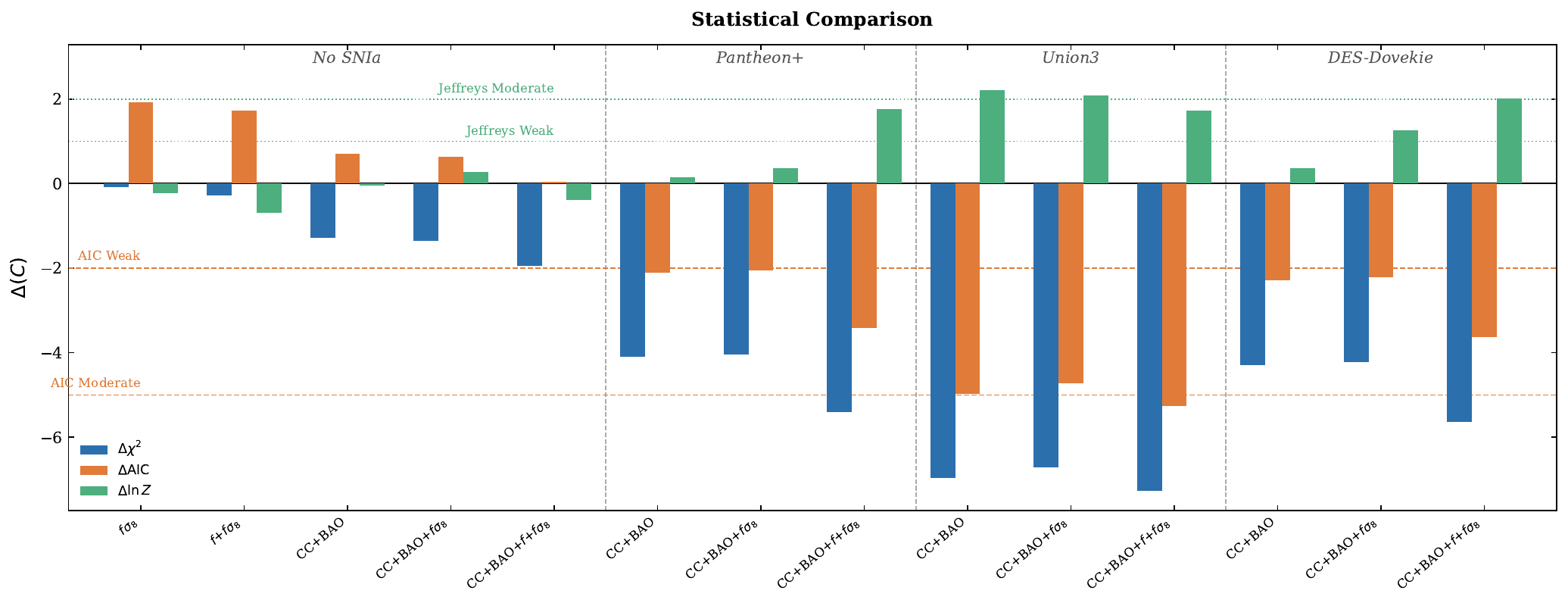}\caption{Graphical
representation of the comparison of the statistical parameters.}%
\label{pla1}%
\end{figure*}

\section{Observational Constraints of the Power-law GUP}

\label{sec4b}

We extend our analysis and for the more general scenario of the power-law
GUP-modified cosmological model. We perform the observational tests only for
three complete combined datasets as presented in Table \ref{tab4}.

From the {\small PP+CC+BAO+}$f${\small +}$f\sigma_{8}$, the analysis of the
MCMC chains reveal $H_{0}=68.1_{-1.7}^{+1.7}$, $~\Omega_{m0}=0.304_{-0.015}%
^{+0.015}$, $\sigma_{8,0}=0.801_{-0.027}^{+0.027}$, $S_{8,0}=0.806_{-0.026}%
^{+0.026}$, $\beta=-0.026_{-0.020}^{+0.027}$ and $\mu=2.1_{-3.3}^{+1.4}$.
Furthermore, by using the U3 SNIa catalogue, that is from the combined data
{\small U3+CC+BAO+}$f${\small +}$f\sigma_{8}$ we determine $H_{0}%
=66.4_{-1.9}^{+1.9}$, $~\Omega_{m0}=0.316_{-0.014}^{+0.014}$, $\sigma
_{8,0}=0.805_{-0.026}^{+0.026}$, $S_{8,0}=0.805_{-0.026}^{+0.026}$,
$\beta=-0.026_{-0.020}^{+0.027}$ and $\mu=4.4_{-4.0}^{+2.2}$. Finally, from
the data set {\small DD+CC+BAO+}$f${\small +}$f\sigma_{8}$ we calculate
$H_{0}=67.7_{-1.6}^{+1.6}$, $~\Omega_{m0}=0.306_{-0.012}^{+0.012}$,
$\sigma_{8,0}=0.801_{-0.025}^{+0.025}$, $S_{8,0}=0.809_{-0.025}^{+0.025}$,
$\beta=-0.024_{-0.005}^{+0.020}$ and $\mu=3.8_{-4.2}^{+2.2}$.

We remark that the value for the quadratic theory, $\mu=2$, is always within
the $68\%$ CI. The power-law GUP theory provides slightly larger values for the
$\Omega_{m0}$, closer to that of the $\Lambda$CDM, and smaller values for
the $\sigma_{8,0}$ from the quadratic model, but with similar
$S_{8,0}$. As far as parameter $\beta$ is concerned, the median values are
indeed negative, however, with larger uncertainties, such that the value
$\beta=0$, to be within the $68\%$ CI when the PP SNIa data are used.

In Table \ref{tab4a}, we summarize the comparison of the statistical
indicators for the power-law GUP-modified model and the $\Lambda$CDM. The
model provides fits with smaller $\chi_{\min}^{2}$ values, including in
comparison with the quadratic GUP model. However, due to the presence of an
additional degree of freedom, it is penalized by both the AIC and the Bayesian evidence.

According to the Akaike scale, there is weak preference for the power-law GUP
model when the combined dataset includes the PP and DD SNIa catalogues, and
strong preference when the U3 catalogue is used. Nevertheless, the Bayesian
evidence shows that the power-law GUP model is statistically equivalent to
$\Lambda$CDM.%

\begin{table*}[tbp] \centering
\caption{Numerical outcomes for the GUP-modified cosmology beyond the quadratic model.}%
\begin{tabular}
[c]{cccccccc}\hline\hline
\textbf{Dataset} & $\mathbf{H}_{0}\mathbf{~}$ & $\mathbf{\Omega}_{m0}$ &
$\mathbf{r}_{drag}$ & $\mathbf{\mu}$ & $\mathbf{\beta}$ & $\mathbf{\sigma
}_{8,0}$ & $\mathbf{S}_{8,0}$\\\hline
\multicolumn{8}{c}{\textbf{With SNIa}}\\
{\small PP+CC+BAO+}$f${\small +}$f\sigma_{8}$ & $68.1_{-1.7}^{+1.7}$ &
$0.304_{-0.015}^{+0.015}$ & $147.2_{-3.5}^{+3.5}$ & $2.1_{-3.3}^{+1.4}$ &
$-0.026_{-0.020}^{+0.027}$ & $0.801_{-0.027}^{+0.027}$ & $0.806_{-0.026}%
^{+0.026}$\\
{\small U3+CC+BAO+}$f${\small +}$f\sigma_{8}$ & $66.4_{-1.9}^{+1.9}$ &
$0.316_{-0.014}^{+0.014}$ & $147.4_{-3.5}^{+3.5}$ & $4.4_{-4.0}^{+2.2}$ &
$-0.037_{-0.011}^{+0.022}$ & $0.805_{-0.026}^{+0.026}$ & $0.826_{-0.028}%
^{+0.028}$\\
{\small DD+CC+BAO+}$f${\small +}$f\sigma_{8}$ & $67.7_{-1.6}^{+1.6}$ &
$0.306_{-0.012}^{+0.012}$ & $147.2_{-3.5}^{+3.1}$ & $3.8_{-4.2}^{+2.2}$ &
$-0.024_{-0.005}^{+0.020}$ & $0.801_{-0.025}^{+0.025}$ & $0.809_{-0.025}%
^{+0.025}$\\\hline\hline
\end{tabular}
\label{tab4}%
\end{table*}%
%

\begin{table*}[tbp] \centering
\caption{Statistical comparison of the power-law GUP-modified model with standard $\Lambda$CDM using Bayesian Evidence and AIC}%
\begin{tabular}
[c]{cccccc}\hline\hline
\textbf{Dataset} & $\Delta\chi_{\min}^{2}$ & $\Delta\left(  AIC\right)  $ &
\textbf{Akaike's scale} & $\Delta\left(  \ln Z\right)  $ & \textbf{Jeffrey's
scale}\\\hline
\multicolumn{6}{c}{\textbf{With SNIa}}\\\hline
{\small PP+CC+BAO+}$f${\small +}$f\sigma_{8}$ & $-5.62$ & $-1.62$ & Weak favor
for GUP & $-0.17$ & Inconclusive\\
{\small U3+CC+BAO+}$f${\small +}$f\sigma_{8}$ & $-9.90$ & $-5.40$ & Strong
favor for GUP & $+0.99$ & Inconclusive\\
{\small DD+CC+BAO+}$f${\small +}$f\sigma_{8}$ & $-7.52$ & $-3.52$ & Weak favor
for GUP & $+0.53$ & Inconclusive\\\hline\hline
\end{tabular}
\label{tab4a}%
\end{table*}%

\section{Conclusions}

\label{sec5}

The quantum corrections terms arising from GUP, due to the existence of a
minimal observable length at the Planck scale, are expressed with the
modification of the canonical algebra and the corresponding Poisson brackets.
In the FLRW cosmological framework, this leads to a modified Raychaudhuri
equation in which nonlinear terms are introduced which affect the dynamical
evolution and modify the physical properties for the standard $\Lambda$CDM
model. Thus, the theory can be described phenomenologically as a one-parameter
dynamical dark energy model. The additional parameter is the deformation
constant, which characterizes the underlying GUP structure. Previously,
background cosmological observations were used to constrain the deformation
parameter and to examine the viability of the GUP-modified model as a dark
energy candidate. In this work, we extended this analysis by employing for the
first time RSD measurements to constrain the deformation parameter through the
evolution of matter perturbations. Such analysis provide tight constraints on
the deformation parameter and allow us to investigate the effects of the GUP
on the evolution of the matter perturbations. 

For the quadratic GUP, by using the closed-form expression for the Hubble
function, we performed a detailed analysis on the cosmological constraints by
using different combinations of the data sets employed in this work. From the
RSD data, we found that there exists a systematic preference for a negative
median value for the deformation parameter $\beta$. The inclusion of the SNIa
data gives tight constraints on the deformation parameter, with the $\Lambda
$CDM limit, i.e. $\beta=0$, to lie within the $95\%$ CI but not at the $68\%$
level. Moreover, the model supports smaller median values for the rest of the
physical parameter $H_{0}$, and $\Omega_{m0}$ in comparison with the obtained
median value for the base model, which is the $\Lambda$CDM. Moreover, as far
as the $\sigma_{8,0}$ and $S_{8,0}$ parameters are concerned, the obtained
values are lower from that of the Planck 2018 and have similar limits with
that of the $\Lambda$CDM. 

By using the AIC and the Bayesian evidence, we investigated if the data sets
have a preference model. For the RSD data combined with the CC and the BAO
observations no statistically significant difference is found between the two
models. However, when the SNIa data are considered in the analysis, the AIC
reveal a weak or strong support in favor of the GUP-modified cosmological
model. In particular when the PP and DD data are used, the support is weak, while
for the U3 catalogue the support is strong. On the other hand, the
application of Jeffrey's scale for the Bayesian evidence, indicate a weak
evidence in favor of the GUP-model, when the U3 and DD catalogues are combined
with the RSD observables. Therefore, these results should be understood as
cosmological limits on the deformation parameter, not as direct evidence for
the existence of a minimal length. 

In Fig. \ref{pla4a} we present a reconstruction plot for the comparison of the
$f\left(  z\right)  $ and $f\sigma_{8}\left(  z\right)  $ data with the
theoretical predictions obtained by the constraints given by Tables \ref{tab1}
and \ref{tab1b} within the credible interval at the $68\%$ level. Finally, in
Fig. \ref{pla5a} we reconstruct the growth index $\gamma\left(  z\right)
=\frac{\ln f\left(  z\right)  }{\ln\Omega\left(  z\right)  }$, where we show a
small deviation from the value $\gamma=\frac{6}{11}$ obtained for the
$\Lambda$CDM.

Furthermore, we considered a power-law GUP model, with introduces the
additional parameter $\mu$, which describes the power of the deformation
function. We employed the same combined data sets, and we found that the
quadratic model, $\mu=2$, is always within the $68\%$ CI. Parameter $\mu$ is
poorly constrained, such that the model to provide to be penalized from the AIC
and the Bayesian evidence.

\begin{figure*}[h]
\centering\includegraphics[width=1\textwidth]{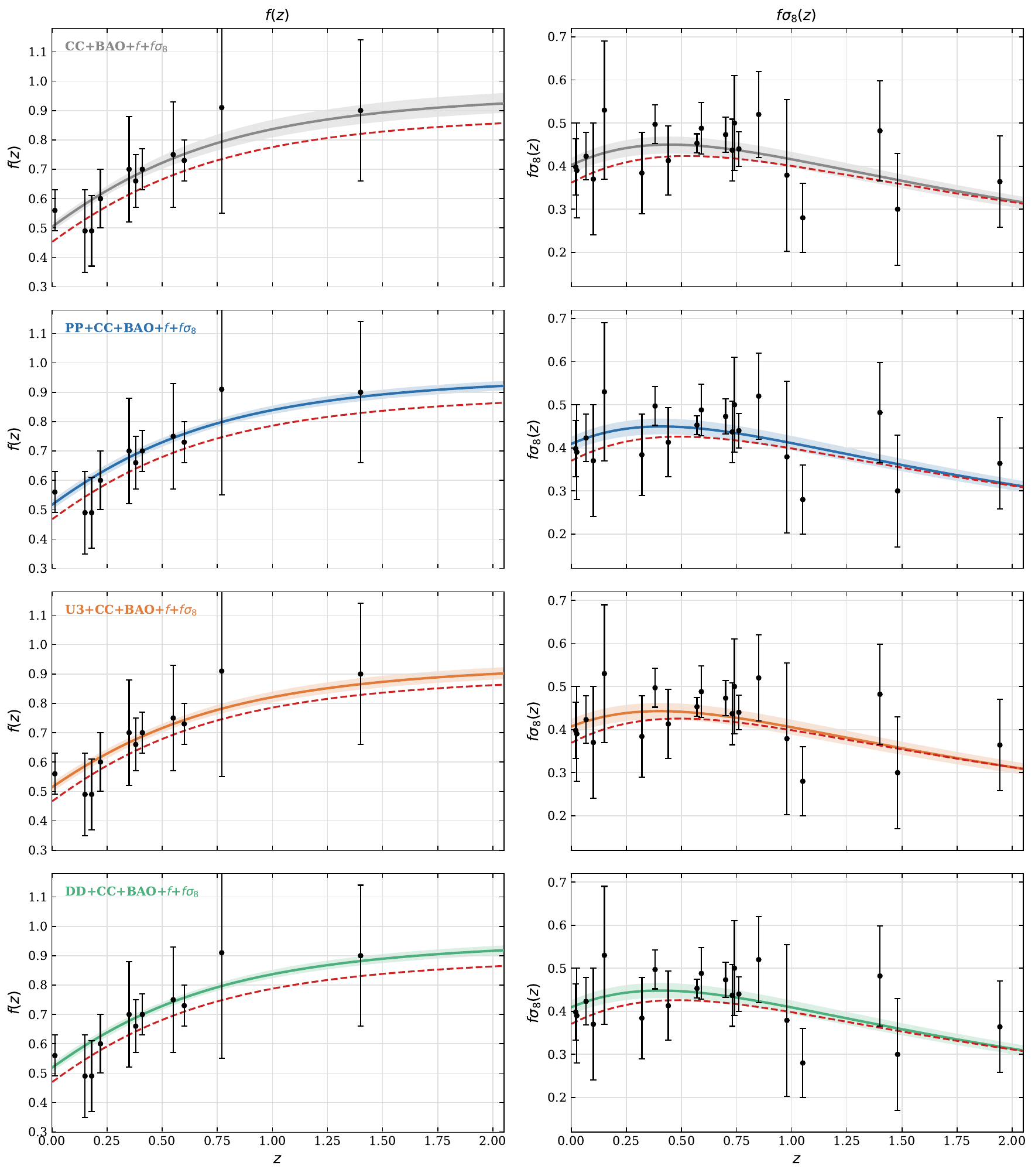}\caption{Reconstruction
of $f\left(  z\right)  $ and $f\sigma_{8}\left(  z\right)  $ for the quadratic
GUP-modified cosmological model for the variables and the 68\% CI as presented
in Tables \ref{tab1} and \ref{tab1b} as obtained from the different data
sets.}%
\label{pla4a}%
\end{figure*}

\begin{figure}[h]
\centering\includegraphics[width=0.5\textwidth]{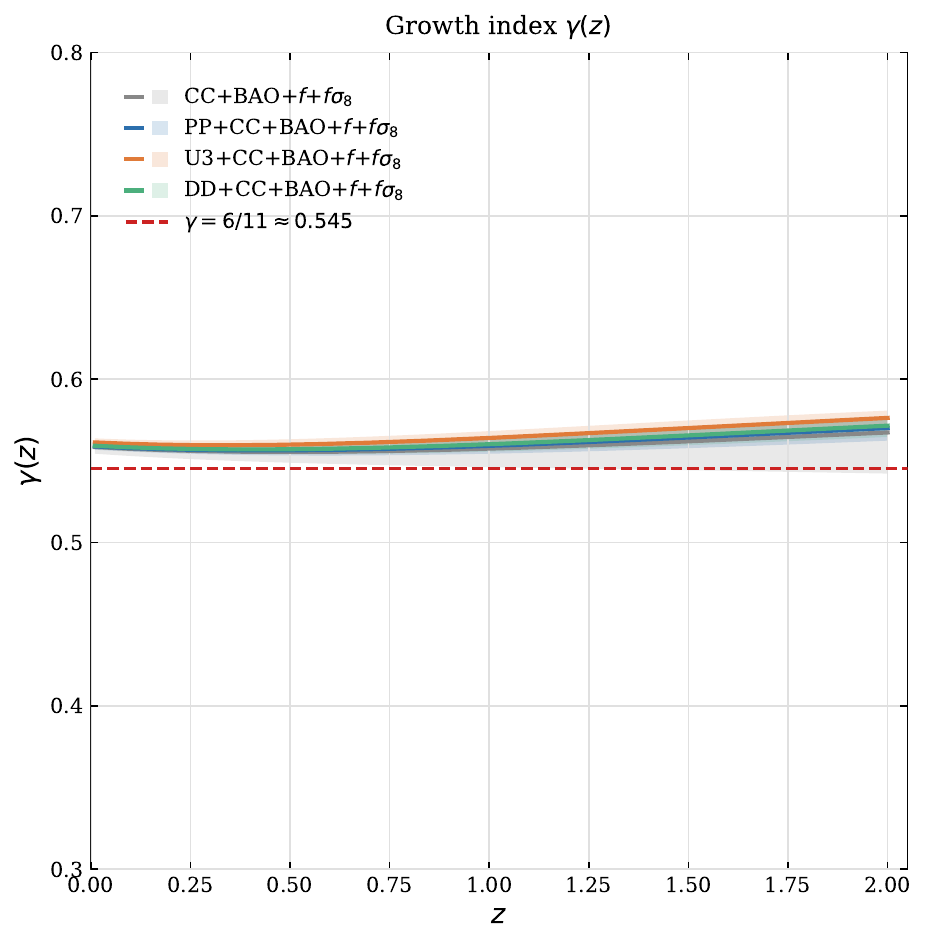}\caption{Evolution of
the growth of index parameter $\gamma\left(  z\right)  =\frac{\ln f\left(
z\right)  }{\ln\Omega\left(  z\right)  }$ for the variables and the 68\% CI as
presented in Tables \ref{tab1} and \ref{tab1b} as obtained from the different
data sets and comparison with the $\Lambda$CDM value $\gamma_{\Lambda
\text{CDM}}=\frac{6}{11}$. }%
\label{pla5a}%
\end{figure}
\begin{acknowledgments}
AP thanks the support of VRIDT through Resoluci\'{o}n VRIDT No. 096/2022 and
Resoluci\'{o}n VRIDT No. 021/2026. Part of this study was supported by
FONDECYT 1240514.
\end{acknowledgments}

\bibliography{biblio}

\end{document}